\documentclass{iopart}
\usepackage{graphicx,amsfonts,amsbsy,amssymb,amsthm}
\newcommand{\be}{\begin{equation}}
\newcommand{\ee}{  \end{equation}}
\newcommand{\ba}{\begin{eqnarray}}
\newcommand{\ea}{  \end{eqnarray}}

\def\={\;=\;} \def\+{\,+\,}

\begin{document}

\title{Delay-time distribution in the scattering of time-narrow wave packets.  (I) }

\author{ Uzy Smilansky}
\address { Department of Physics of Complex Systems,
 Weizmann Institute of Science, Rehovot 7610001, Israel.}

 \date{\today}

\begin{abstract}
 This is the first of two subsequent publications where the probability distribution of delay-times in scattering of wave packets is discussed. The probability distribution is expressed in terms of the on-shell scattering matrix, the dispersion relation of the scattered beam and the wave packet envelope. In the monochromatic limit (poor time resolution) the mean delay-time coincides with the expression derived by Eisenbud and Wigner and generalized by Smith more than half a century ago. In the opposite limit, and within the semi-classical approximation, the resulting distribution coincides with the result obtained using classical mechanics or geometrical optics. The general expression interpolates smoothly between the two extremes. An application for the scattering of electromagnetic waves in networks of RF transmission lines will be discussed in the next paper to illustrate the method in an experimentally relevant context.

\end{abstract}
\maketitle
\section{Introduction}
This study is motivated by the new experimental research directions opened by the availability of ultra-fast radiation pulses. The most prominent example are the light pulses (wave-packets) of duration comparable to one cycle of the carrier electro-magnetic field \cite{Ultrafast1, Ultrafast2,Hassan,Anlage}. When such a pulse is scattered on a complex target, the field is trapped for some time, and emerges as a time broadened  pulse, whose shape reflects the distribution of delay-times. Intuitively, the term  delay-time stands for the differences between the time it would take  the scattered particle (or wave) to transit through the domain of interaction and the time of transit when the interaction is switched off.  Note that the delay-time can also assume negative values, as  happens e.g., in scattering of quantum particles by attractive potentials. The difficulty in the above definition stems from the fact that it is not always clear how to disentangle the net effect of the interaction from the complete evolution.  Classical scattering theory provides a precise definition of the net delay which overcomes this difficulty. The classical definition of the delay time distribution will be discussed briefly in the beginning of the next section. However, the classical definition cannot be simply transplanted to the quantum (or wave) scattering for various reasons, the most obvious is that the classical delay-time depends on the precise value of the energy, which might lead to conflicts with the uncertainty principle.

Quantum scattering theory circumvents this problem since it uses the scattering operator
\begin{equation}S=\lim_{t\rightarrow \infty} e^{\frac{i}{\hbar}H_0 \frac{t}{2}}e^{-\frac{i}{\hbar}H t }e^{\frac{i}{\hbar}H_0\frac{t}{2}}\ .
\label{scatop}
\end{equation}
It measures the difference  between the  evolution induced by the full Hamiltonian $H$ and the free evolution induced by the asymptotic Hamiltonian $H_0$. This is therefore the natural tool for defining the quantum delay-time distribution, and this is done in the next section in terms of  the scattering matrix $S(E)$ - the restriction of the scattering operator (\ref {scatop}) to an  energy shell. The scattering process describes the transition from an initial (incoming)  state which is an eigenstate $|i \rangle$  of the free Hamiltonian $H_0$  to a final (outgoing)  eigenstate $|f \rangle$. The matrix elements $S_{f,i}(E)=\langle f|S(E)|i \rangle $ give the scattering probability amplitude for the $i\rightarrow f$ transition. The derivation of the delay-time distribution  presented here is rooted in the scattering matrix formalism,  and it will be shown to  approach   the intuitive classical limit.

The delay-time distribution depends crucially on the dispersion relation $E(k)$ of the waves under consideration, and on the envelope of the incoming wave-packets. Gaussian wave packets of electromagnetic radiation with $E(k)=ck$ propagate without distortion (dispersion) as long as the material is uniform. In contrast, free quantum wave packets with $E(k)=\frac{\hbar^2}{2m}k^2$ remain Gaussian but their variance increases quadratically in time. For this reason  these two cases will be discussed separately. Absorption and other dissipative processes will be neglected for EM waves,  however, inelastic scattering  will be discussed in the context of quantum scattering.

The definition and the properties of the time delay in quantum scattering were discussed by several authors in the past. The pioneers were E. P. Wigner and his followers \cite{Eisenbud,Bohm,Wigner,Smith}. They did not attempt to compute the distribution of time -delays. Rather, they computed the {\emph mean} delay time - the first moment of the delay-time distribution - in terms of the scattering matrix at a prescribed energy. However, Wigner's derivation explicitly assumes that the incoming radiation is almost monochromatic and the uncertainty principle restricts it to pulses of long duration.  The present theory recovers Wigner's expression of the mean delay time in the monochromatic limit, where the incoming pulses are very broad in time. It also yields the second moment, which diverges  as the monochromatic limit is approached.

The delay-time distribution is expressed here in terms of the auto-correlation of the $S$ matrix,
\begin{equation}
P_{f,i}(\tau) \sim \int{\rm d}\epsilon\  e^{\frac{i}{\hbar} \epsilon \tau}\left \langle S_{f,i}(E+\frac{\epsilon}{2})
S^{\star}_{f,i}(E-\frac{\epsilon}{2})\right\rangle _E \ ,
\label{firstdef}
\end{equation}
where $\tau$ is the delay time,  the transition for which the delay is measured is $i\ \rightarrow\ f$ and the triangular brackets stands for an average over the incoming energy. The precise way in which  the envelope of the incoming wave-packet is reflected in the averaging will be discussed in the proceeding sections.  A similar definition was first introduced by V.L. Lyuboshits back in 1983 \cite{Lyuboshits}, for the analysis of overlapping resonances in reactions between nuclei. A few years later  it was independently proposed by R. Bl$\ddot{{\rm u}}$mel and the present author in the semi-classical study of chaotic scattering \cite{blumel}. In both works the energy averaging was not defined in a precise way. The present paper expands upon these studies, and provides a uniform framework where previously proposed approximations and limits can be derived within the same conceptual structure. See \cite{{Hartman},{Buttiker-Land},{Buttiker},{Pollak Miller},{Sokolovsky},{Land-Martin},{Yamada},{Texier}}, and references cited therein.

The term "delay-time distribution" appears frequently in the literature. It usually refers to a distribution defined over an ensemble of systems (e.g.,  random matrix ensembles) and the quantity computed for each system is Wigner's mean delay-time. The quantity presented here is different - it is the distribution of delay times for a given scattering systems.

The second  paper in this series will illustrate the method  by applying it to the distribution of delay-times for scattering through  metric graphs. In this case, an exact multiple scattering formalism can be used to express the scattering matrix, and this in turn allows a derivation of the asymptotic form of the delay-time distribution  in terms of the graph connectivity and  metrics.  The propagation of waves in graphs can be simulated by the propagation of RF waves through networks of transmission lines.  Such an experiment which was recently performed by S. Anlage and coworkers \cite{Anlage}, and it stimulated the present work.

\section{Scattering approach to delay-time distribution}
The delay-time in scattering is a well defined concept in classical mechanics. Consider e.g., a particle which is scattered by a non-spherical, time independent potential $V(r,\theta,\phi)$ which decays faster than $r^{-1}$ as $r \rightarrow \infty$.  Denote the particle phase-space  trajectory by $({\bf r}(t),{\bf p}(t))$.  At $t=-T$,  with $T$ very large but finite, the incoming momentum approaches a constant ${\bf p}_i $ which is directed along the unit vector ${\bf \Omega}_i$ specified by the point $\Omega_i=(\theta_i,\phi_i)$ on the unit sphere. It carries the angular momentum  ${\bf l}_i={\bf r}(-T) \times{\bf p}_i $. After the interaction, the outgoing trajectory at $t\rightarrow +T$ is characterized by an outgoing momentum  ${\bf p}_f $ pointing in the direction of the unit vector ${\bf \Omega}_f$, with   $\Omega_f=(\theta_f,\phi_f)$. Energy conservation dictates $|{\bf p}_i |=|{\bf p}_f |$, but the final angular momentum ${\bf l}_f={\bf r}(+T) \times{\bf p}_f $ does not equal ${\bf l}_i$  when the potential is not spherically symmetric. The scattering event is thus defined as the transition $\Omega_i \rightarrow  \Omega_f$ at energy $E=\frac{|{\bf p}_i|^2}{2m}$. The time it takes to make this transition is $2T$ which can take any large but otherwise arbitrary value. However, the  net delay time is the difference between $2T$ and $2T_0$ - the time it would have taken to make the transition along unperturbed trajectories. Both $T$ and $T_0$  are divergent quantities. Their difference, as $T\rightarrow \infty$ approaches a constant and this is the proper definition of a delay time. To put it in a precise form, consider the (regularized) action integral along the trajectory which is specified by the incoming energy and the incoming and outgoing directions \cite {leshouches}.
\begin{eqnarray}
\label{action}
\hspace{-10mm} \Phi(\Omega_f,\Omega_i;E)&=&-\int_{-T}^{+T}{\rm d}t\  \left ({\bf r}(t)\cdot {\bf \dot{p}}(t)\right ) \\
 &=&\ \ \ \int_{-T}^{+T}{\rm d}t\  \left ({\bf \dot{r}}(t)\cdot {\bf p}(t)\right ) - \left [\left({\bf p}_f\cdot {\bf r}(+T)\right )-\left ({\bf p}_i\cdot  {\bf r}(-T)\right )\right ] \nonumber ,
 \end{eqnarray}
 where a dot above stands for time derivative. The integral in the lower line of (\ref {action}) is the standard action integral which clearly diverges when $T\rightarrow \infty$. The second term in the same line is the boundary term which also diverges. However, their difference (the integral in the first line) gets no contribution from large times  since as $T\rightarrow\infty$, ${\bf \dot{ p}}  \rightarrow 0$ sufficiently rapidly so the integral is finite.  Taking the  derivative of $\Phi$  with respect to $E$ we get
 \begin{equation}
 \tau_{\Omega_i ,\Omega_f}(E) =\frac{\partial \Phi(\Omega_f,\Omega_i;E)}{\partial E} = 2T-\frac{1}{v}(|{\bf r}(-T)|+|{\bf r}(T)|)\ ,
 \label{tauclass}
\end{equation}
where $v=\sqrt{\frac{2E}{m}}$ is the asymptotic velocity. The second term equals  $2T_0$ the time it would have taken to traverse freely the distance  $|{\bf r}(-T)|+ |{\bf r}(T)|$ . Thus, in the limit $T\rightarrow \infty$, (\ref {tauclass}) is the classical net delay-time associated with the considered trajectory.

When the dynamics in the scattering process is complex, there are many trajectories which satisfy the same boundary conditions, that is, scattering from $\Omega_i$ to $\Omega_f$ at energy $E$. Denote this set of trajectories by $\mathcal{A}(\Omega_f,\Omega_i;E)$. Classical mechanics assigns to each trajectory $\alpha\in \mathcal{A} $ a probability $p_{\alpha}$ to occur,  and therefore, one can define the classical delay-time distribution as
\begin{equation}
P^{class}_{\Omega_f,\Omega_i}(\tau;E)=\sum_{\alpha\in \mathcal{A}(\Omega_f,\Omega_i;E) }p_{\alpha}\delta(\tau-\tau_{\alpha}).
\label{distclass1}
\end{equation}

The concepts defined or derived above for elastic scattering,  can be easily generalized to include most types of reactions as long as the ingoing and outgoing channels involve two particles (no break-up reactions). We shall return to the classical concepts  in section (\ref {qm}) where we shall derive the semi-classical expression for the delay-time distribution. These concepts can be as well transcribed to geometric optic, by considering $V({\bf r})$ as the local refractive index, and replacing the action by the optical length.

\

The net delay-time as defined in (\ref {distclass1}) is a classical quantity which cannot be directly used in the discussion of wave-scattering since it provides precise information on a time interval at a specified energy - disregarding the uncertainty principle.  The quantum (wave) dynamical definition of the delay-time distribution proposed here, naturally overcomes this difficulty. To write it down explicitly, denote the (Unitary) scattering matrix by $S_{f,i}(E)$. Consider an incoming wave-packet incoming from the reaction channel $i$, and with an envelope function $\omega(E)$ normalized by  $\int_{0}^{\infty}{\rm d}E \omega^2(E) =1$. The field intensity (number of particles) is measured in the reaction channel $f$. The  delay-time distribution for the $i \rightarrow f$ transition will be shown to be
\begin{equation}
P_{f,i}(\tau) = \frac{1}{2\pi\hbar}\left | \int_0^{\infty} {\rm d}E\ \omega(E) S_{f,i}(E) e^{-\frac{i}{\hbar}E\tau} \right |^2\ .
\label{poftfirst}
\end{equation}
The  unitarity of $S(E)$  and the normalization of $\omega(E)$ guarantee that
\begin{equation}
\int_{-\infty}^{\infty} {\rm d}\tau  \sum_f P_{f,i}(\tau) = 1.
\end{equation}
Thus $P_{f,i}(\tau)$ is a positive and properly normalized function which satisfies the necessary conditions that a probability density must satisfy.  The following argument (which is similar to the one used in \cite {Lyuboshits}) will show that it provides a natural choice for the delay time distribution. A well known expression for the scattering matrix  \cite {Feshbach}
\begin{equation}
S_{f,i}(E)= S^{(P)}_{f,i}(E)-i\sum_{\mu}\frac{g_{ f,\mu}g_{\mu,i}}{E-E_{\mu}+\frac{i}{2}\Gamma_{\mu}}
\end{equation}
gives the $S$ matrix in terms of its poles in the lower part of the complex $E$ plane. The partial widths $g_{\mu,i}$ must satisfy $\sum_k g_{\mu,k}g_{\mu',k}=\delta_{\mu,\mu'}\Gamma_{\mu}$ to render $S$  unitary. $S^{(P)}_{f,i}(E)$ is the contribution from "prompt" (or "direct") processes which are characterized by a slowly varying $S^{(P)}_{f,i}(E)$, and hence cause no appreciable delay. To simplify the presentation, assume that the partial widths $g_{\mu,k}$ are constants. In this approximation,
\begin{eqnarray}
\label{probdist0}
\hspace{-10mm}
P_{f,i}(\tau)&=&\frac{1}{ 2\pi\hbar } \left \{ \sum_{\mu}|\omega(E_{\mu}-\frac{i}{2}\Gamma_{\mu})|^2|g_{\mu,i}g_{\mu,f}|^2 e^{-\frac{1}{\hbar}\Gamma_{\mu}\tau}\right . \\   \hspace{-10mm}
&+&\left . \sum_{\mu\ne \mu' }G_{i,f}(\mu,\mu')e^{\frac{i}{\hbar}(E_{\mu}-E_{\mu'})\tau}
 e^{-\frac{1}{2 \hbar}(\Gamma_{\mu}+\Gamma_{\mu'})\tau} \right \}\nonumber \  ,
\nonumber
\end{eqnarray}
where, $ G_{i,f}(\mu,\mu') =\omega(E_{\mu}-\frac{i}{2}\Gamma_{\mu})\omega(E_{\mu'}+\frac{i}{2}\Gamma_{\mu'})g_{\mu,i}g_{\mu,f}g_{\mu',i}g_{\mu',f}$.

\

\noindent The first line in (\ref{probdist0}) is the leading term. It is a sum over all resonances within the effective support of $\omega(E)$, each given  the relative strength  to be excited in the $i\rightarrow f$ transition, and each decaying exponentially with its typical time constant $\frac {\hbar}{\Gamma_{\mu}}$. This conforms with the intuitive expectation from the delay-time distribution.
The double sum  in (\ref{probdist0}) is composed of oscillatory terms with phases of order of the mean spacing between successive resonances  $(\Delta E)$ divided by $\hbar$. Thus the double sum is expected to be negligible for times larger than the Heisenberg time $\frac{\hbar}{\langle \Delta E\rangle}$.

Finally, the form (\ref {poftfirst}) lends itself naturally to be written as the Fourier transform of the autocorrelation function of $S_{f,i}(E)$ as in (\ref{firstdef}). Thus, the decay-time distribution is the so-called form-factor of $S_{f,i}(E)$. This will be exploited in the next section.

The delay-time distribution depends on the dispersion relation, and requires somewhat different treatments. The simpler case of linear dispersion will be treated in the first subsection. The modifications needed for quantum scattering will be presented in the second subsection, together with a  detailed semi-classical analysis.

\subsection{Scattering of Electromagnetic waves}
\label{em}
Consider the scattering of a train of electromagnetic wave packets on a complex set of perfectly reflecting surfaces confined in a finite  volume.  Typical examples consist of e.g., a scatterer formed by a number of reflecting surfaces, or a finite, arbitrarily shaped block of material with spatially dependent refractive index.  The electromagnetic field will be described as a  classical field, and polarization will be ignored for the sake of simplicity.

Assume that the incoming beam is directed along the unit vector ${\bf \Omega }_i$  represented by a point $ \Omega_i =(\theta_i,\phi_i)$ on the unit sphere. The incoming wave function is a  wave packet (pulse)  expressed as a superposition of the plane waves $ e^{ik({\bf r}.{\bf \Omega}_i)}$  multiplied by the time dependent phase $e^{-i ks }$, where $k$ is the wave number and $s=ct$ is the path length traversed during the time $t$, and $c$ is the speed of light.
The wave packet is defined by an  envelope function $\omega(k)$ which is assumed to be non-negative, to posses a single maximum at the carrier wave-number $k_0$, and to be normalized $\int_0^{\infty}\omega^2(k){\rm d}k=1$. It is also convenient to associate with $\omega(k)$ a parameter $\sigma$ which characterizes the extent of the $k$ domain where $\omega(k)$ is large. Otherwise, no further restriction is imposed on the pulse shape. With this envelope function the time dependent incoming field  is
 \begin{equation}
\mathcal{E}_{in}( {\bf r},t)=  \int_{0}^{\infty}{\rm d}k\  \omega(k)
e^{ik( {\bf r} .{\bf \Omega}_i)}e^{-i ks } ,
\label{initial}
\end{equation}
 It propagates in the  direction  ${\bf\Omega}_i $,   the  envelope is time independent and it is constant in the transversal directions. This form is expected to approximate  the spacial and temporal structures of the experimentally produced short laser pulses.

The scattering process is completely characterized by the scattering matrix $S_{  \Omega_f, \Omega_i}(k)$ which provides the amplitude for an incoming wave in the direction ${  \Omega_i}$ with wave number $k$  to scatter into the direction ${  \Omega_f}$. The scattering matrix is unitary and satisfies for all $k$
 $$ \int_{S^2} {\rm d}\Omega\
  S_{  \Omega_f, \Omega }(k)S^{\dagger}_{  \Omega, \Omega_i}(k)= \delta(\Omega_f-\Omega_i)\ . $$

Consider now a wave-packet approaching in the ${\Omega_i}$ direction. The field  intensity observed in a small solid angle ${\rm d}\Omega$  about the directed  ${\Omega_f}$ with a delay-time $s=ct$ is expressed as
\begin{equation}
\hspace{-15mm}
I_{{  \Omega_f},{  \Omega_i}}(s){\rm d}\Omega =|\mathcal{E}_{out}(s)|^2 {\rm d}\Omega =  \frac{1}{2\pi}\  |\int_{0}^{\infty} {\rm d}k   \ \omega(k) S_{  \Omega_f, \Omega_i}(k) e^{-iks}\ \ |^2 {\rm d}\Omega
\label{poftem1}
\end{equation}
 The normalization of the wave packet envelope $\omega(k)$, and the unitarity of the $S(k)$ matrix at each $k$, imply that for any incoming direction,
\begin{equation}
\int_{S^2} {\rm d}{  \Omega_f} \int_{-\infty}^{\infty}I_{{ \Omega_f},{ \Omega_i}}(s) {\rm d} s  =1.
\label{normalization}
\end{equation}
 The function $I_{{\Omega_f},{\Omega_i}}(s) $ can also be considered as the probability distribution of the delay-times.  It is clearly a non-negative function of $s$ and it is properly normalized (\ref {normalization}). Hence, it satisfies the elementary requirement for a probability density, and it  will be referred to in the remainder of the article as the delay-time probability distribution and will be denoted by $P_{{\Omega_f},{\Omega_i}}(s)$ - to match with the notation in the quantum mechanical treatment.

 Writing (\ref {poftem1}) as a product of two integrals in $k$ and $k'$ and defining $\xi=\frac{1}{2}(k+k')$ and $\eta=(k-k')$ we get
\begin{eqnarray}
\hspace{-5mm} P_{{\Omega_f},{\Omega_i}}(s)&=&\frac{1}{2\pi} \int_{0}^{\infty}{\rm d}\xi \int_{-2\xi}^{2\xi}{\rm d}\eta\ e^{-i \eta  s} \omega(\xi+\frac{\eta}{2})\omega(\xi-\frac{\eta}{2})\times
  \nonumber  \\
& &    S_{{\Omega_f},{\Omega_i}}(\xi+\frac{\eta}{2})\ S^{\star}_{{\Omega_f},{\Omega_i}}(\xi-\frac{\eta}{2})  .
\label{poftem2}
\end{eqnarray}
This expression  is now used to show that in  the monochromatic limit ($\sigma \rightarrow 0 $), the mean delay time coincides with the Wigner-Smith expression. For this purpose, write the mean delay-time during scattering from $\Omega_i$ to $\Omega_f$,
$$\langle s\rangle_{{\Omega_f},{\Omega_i}}=\int_{-\infty}^{\infty} {\rm d}s \ s\  P_{{\Omega_f},{\Omega_i}}(s)\ .$$
$\langle s\rangle_{{\Omega_f},{\Omega_i}}$ is computed from (\ref {poftem2}) by first introducing $ i \frac{\partial\ }{\partial \eta}$ in front of  $e^{-i\eta s}$ in the upper line of (\ref {poftem2}), and integrating over the entire range of $s$. The $\eta$ integral is  carried out by parts, followed by the ${\rm d}s$ integral which results in $2\pi \delta (\eta)$.  Finally,
\begin{eqnarray}
\hspace{-22mm}
\langle s\rangle_{{\Omega_f},{\Omega_i}}=
i\ \int_{0}^{\infty}{\rm d}\xi
  \frac{\partial\ }{\partial \eta}\left[\omega(\xi+\frac{\eta}{2})\omega(\xi-\frac{\eta}{2}) S_{{\Omega_f},{\Omega_i}}(\xi+\frac{\eta}{2})\ S^{\star}_{{\Omega_f},{\Omega_i}}(\xi-\frac{\eta}{2})\right ]_{\eta=0}\ .
 \label{tauem}
\end {eqnarray}
 One should bear in mind that $\omega^2(x)$ has a maximum at $k_0$ and it turns into a $\delta(k-k_0)$ function when one approaches the monochromatic limit, $\sigma \rightarrow 0$. Therefore,
\begin{equation}
 \langle s(k_0)\rangle_{{{\Omega_f},{\Omega_i}}}=
 \mathcal{I}{\it m}\left[ S_{{\Omega_f},{\Omega_i}}(k_0 ) \frac{\partial\ }{\partial k}\ S^{ \dag}_{{\Omega_i},{\Omega_f}}(k_0 )\right ] \ .
\end {equation}
Averaging over the initial directions and integrating over the final directions one gets
$$\langle s (k_0)\rangle=\frac{1}{4\pi} \mathcal{I}{\it m}\left[{\rm tr} \left (S(k_0)\frac{\partial\ }{\partial k}\ S^{ \dag}(k_0)\right )\right ]\ . $$
This coincides with the standard form of Smith's version of Wigner's mean delay time \cite {Smith} for elastic scattering from non spherically symmetric scatterers.

It is both customary and  convenient to use for the wave-packet envelope function a square normalized Gaussian
\begin{equation}
\omega(k-k_0)= A(k_0) \left (\frac{2}{\pi\sigma^2}\right )^{\frac{1}{4}}e^{-\frac{(k-k_0)^2}{\sigma^2}}\ ,
\label{gaussian}
\end{equation}
where $A(k_0)$ is a normalization constant (which is necessary since the integration is over the half-line).  All the known techniques to produce coherent ultra-fast pulses require that the width exceeds the wavelength $\frac{1}{k_0}$ of the carrier wave, hence $ k_0 > \sigma$.
Subject to  $k_0> q\sigma$ with arbitrary real $\ q >2$, one can substitute $A(k_0)=1$ and extend  the $\xi$ and $\eta$ integrations in (\ref {poftem2}) to the entire plane,  introducing an error of order $e^{-q^2}$ which can be made smaller than a fraction of a percent - quite  sufficient for most applications. As a result,
\begin{eqnarray}
\hspace{-14mm} P_{{\Omega_f},{\Omega_i}}(s)&\approx &\frac{1}{2\pi}  \int_{-\infty}^{\infty}{\rm d}\eta\ e^{-i \eta  s}
e^{-\frac{ \eta^2}{2 \sigma^2}} \times
  \nonumber  \\
& &\left \{ \sqrt{ \frac{2}{\pi\sigma}} \int_{-\infty}^{\infty}{\rm d}\xi\ e^{-\frac{2(\xi-k_0)^2}{\sigma^2}} S_{{\Omega_f},{\Omega_i}}(\xi+\frac{\eta}{2})\ S^{\star}_{{\Omega_f},{\Omega_i}}(\xi-\frac{\eta}{2})\right \}  .
\label{poftem2prime}
\end{eqnarray}
The term enclosed in the curly brackets is an autocorrelation function of the $S$ matrix element. It is computed with a Gaussian envelope which gets most of its weight from the spectral interval $(k_0-\sigma,k_0+\sigma)$. $P_{{\Omega_f},{\Omega_i}}(s)$ is the Fourier transform of this autocorrelation  which is restricted (again by a Gaussian weight)  to the spectral range $(-\sigma,\sigma)$. Thus it can only provide a temporal resolution of order $\delta t =\frac{\delta s}{c}= \frac{1}{c\sigma}$, as expected.

Starting with (\ref{poftem2prime}) one can compute the second moment of the delay-time distribution in the monochromatic limit. Inserting $-\frac{\partial^2\ }{\partial \eta^2}$ in the ${\rm d}\eta$ integral, and partial integration leads to
\begin{eqnarray}
\hspace{-25mm}
\langle s^2\rangle_{{\Omega_f},{\Omega_i}}=
-\ \int_{0}^{\infty}{\rm d}\xi
  \frac{\partial^2\ }{\partial \eta^2}\left[\omega(\xi+\frac{\eta}{2})\omega(\xi-\frac{\eta}{2}) S_{{\Omega_f},{\Omega_i}}(\xi+\frac{\eta}{2})\ S^{\star}_{{\Omega_f},{\Omega_i}}(\xi-\frac{\eta}{2})\right ]_{\eta=0}\ .
 \label{tauvar}
\end {eqnarray}
For $\sigma$ very small but not vanishing, one gets
\begin{equation}
\hspace{-25mm}
\langle s^2\rangle_{{\Omega_f},{\Omega_i}}= \frac{1}{\sigma^2}|S_{{\Omega_f},{\Omega_i}}(k_0)|^2 -
\frac{1}{2} \mathcal{R}e \left [S^{\star}_{{\Omega_f},{\Omega_i}}(k_0)\frac{ \partial^2 } {\partial k^2}S_{{\Omega_f},{\Omega_i}}(k_0)-|\frac{\partial }{\partial k}S_{{\Omega_f},{\Omega_i}}(k_0)|^2 \right ] .
\end{equation}
In deriving the above equation it was assumed that $\sigma$ is sufficiently small so that the mean value of the scattering matrix and its derivatives over a $\xi$ interval of size $\sigma$ around $k_0$ can be replace as the value at $k_0$. The resulting expression shows that the second moment is inversely proportional to the band-width of the incoming wave-packet, and therefore diverges in the monochromatic limit. This fact emphasizes the utmost importance of computing delay-time distributions taking into account the wave-packets envelope, a feature which was not touched upon in the previous discussions of this topic.

The semi-classical limit ($k_0\rightarrow \infty$) of (\ref{poftem2}) is obtained by writing the semi-classical expression for the $S$ matrix while remaining in the broad-band limit by keeping ${\sigma}=\frac{1}{c}k_0$. In order to avoid duplications, this will be deferred to the next subsection where quantum mechanical scattering will be discussed.  It will be shown there that in this limit $P_{{\Omega_f},{\Omega_i}}(s)$ approaches the classical distribution of the delay times defined in   (\ref {tauclass}).

\subsection {Quantum mechanical scattering}
\label{qm}
Scattering of  particles on molecular or atomic targets usually involve not only the degrees of freedom ${\bf r}$ and their conjugate momenta ${\bf p}$ of the relative motion which were discussed previously, but also the excitation of internal degrees of freedom such as e.g., rotational or vibrational modes \cite {collisiontheory}. Such excitations are distinguished by quantum numbers ${\bf n}=(n_1,\cdots ,n_{F})$ where $F$ is the number of internal freedoms, and the quantum numbers $n_j$ can be considered as the quantized versions  of classical periodic modes described in terms of action-angle variables $(I_j,\varphi_j)$ with $n_j= I_j/\hbar$.
\cite{Miller}. Assuming that the interaction is of finite range, then, away from the interaction region, the asymptotic stationary wave-functions are products of plane-waves in the relative degrees of freedom  and internal eigenfunctions  with quantum numbers ${\bf n}$. The scattering process transforms an incoming asymptotic state $ e^{\frac{i}{\hbar}({\bf p}_i .{\bf r})}\phi_{{\bf n}_i}  $ to a superposition of outgoing states $ e^{\frac{i}{\hbar}({\bf p}_f .{\bf r})}\phi_{{\bf n}_f} $. Conservation of energy requires that
\begin{equation}
\frac{|{\bf p}_f|^2}{2m} + \epsilon_{{\bf n}_f} \ =\ E\ = \frac{|{\bf p}_i|^2}{2m} + \epsilon_{{\bf n}_i}\ ,
\end{equation}
for each outgoing channel where  $\epsilon_{\bf n }$ denote the internal excitation spectrum and $m$ is the reduced mass.

The incoming  wave packet is a superposition of the incoming waves  multiplied by the time dependent phase $e^{\frac{i}{\hbar}(\frac{|{\bf p}_i|^2}{2m}+\epsilon_i)t} $.
Using the same normalized envelope function  as in the previous section,  the time dependent incoming wave-packet is
 \begin{equation}
\psi_{in}({\bf r},t)=  \int_{0}^{\infty}{\rm d}k\  \omega(k-k_0)
e^{ i (k (\Omega_i.{\bf r}) - \frac{\hbar}{2m} k^2 t)} e^{-\frac{i}{\hbar}\epsilon_i t} \ \phi_{{\bf n}_i}\ ,
\end{equation}
where $k=\frac{|{\bf p}_i|}{\hbar}$ and $k_0$ is the mean (carrier) wave number.
Here is where the quadratic dispersion leaves its mark: A freely  propagating Gaussian wave packet undergoes dispersion (broadening), and its effective spacial width $\frac{2}{\sigma}$ at $t=0$    grows after time $t$ to $ \frac{2}{\sigma}  \sqrt{1+ (\frac{\sigma^2 \hbar} {2 m}\ t)^2} $.  The quadratic term starts to dominate  when $\frac{1}{2} (\frac{\sigma}{k_0})^2  (k_0 \delta s) \sim 1$ where $\delta s$ is the path traversed in time $t$. Since $(k_0\delta s)$ equals the number of wavelengths in $\delta s$, the dispersion correction can be neglected for macroscopic $\delta s$  only if the wave-packet is sufficiently monochromatic. Similarly, the measured delay time probability depends on the distance of the target from the detector. This intrinsic difficulty is to find its expression in the theory derived below which is also valid outside the strict monochromatic limit.

The scattering process is completely characterized by the scattering matrix $S_{{\bf n}_f,{\bf n}_f}({\Omega_f} ,{\Omega_i};k)$ which provides the amplitude for an incoming wave in the direction ${\Omega_i}$ with kinetic energy $\frac{\hbar ^2}{2m} k^2$  and the internal system in the eigenstate $\phi_{\bf n_i}$    to scatter to the direction ${\Omega_f}$ leaving the internal system in the eigenstate  $\phi_{\bf n_f}$. The directional information can also be expressed in terms of the components of the angular momentum  $(l_{\theta},l_{\phi})$  which are canonically conjugate to the angle variables $(\theta,\phi)$. To simplify the notation, the arrays ${\bf n}$ of quantum numbers is extended to include the angular momenta quantum numbers.

Consider now a scattering event   which involves the transition from ${\bf n_i}$ to ${\bf n_f}$.  Define,
\begin{equation}
P_{{\bf n_f},{\bf n_i}}(\tau) \doteq \frac{\hbar}{2\pi m}\  |\int_{0}^{\infty} {\rm d}k  \sqrt{k}\ \omega(k-k_{0})S_{{\bf n_f},{\bf n_i}}(k)e^{-\frac{i}{\hbar}E(k)\tau}\ \ |^2
\label{poft1}
\end{equation}
where $E(k)=\frac{\hbar^2}{2m}k^2+\epsilon_i$. By virtue of the normalization of the wave packet envelope $\omega(k)$, and the unitarity of the $S(k)$ matrix at each $k$, one gets in complete analogy to (\ref {normalization})
\begin{equation}
\sum_{{\bf n_f}}\int_{-\infty}^{\infty}P_{{\bf n_f},{\bf n_i}}(\tau) {\rm d} \tau  =1.
\end{equation}
The $\sqrt{k}$ factor  in the integrand of (\ref {poft1}) is necessary for the validity of the identity above.

 Using the Gaussian form of the envelope function and changing the integration variables to $\xi=\frac{1}{2}(k+k')$ and $\eta= (k-k')$, (\ref{poft1}) reads now,
\begin{eqnarray}
\hspace{-15mm} P_{{\bf n_f},{\bf n_i}}(\tau)&=&\frac{\hbar}{2\pi m} \sqrt{\frac{2}{\pi \sigma^2}}\int_{0}^{\infty}{\rm d}\xi\int_{-2\xi}^{2\xi}{\rm d}\eta \sqrt{\xi^2-(\frac{\eta}{2})^2}\ \times  \nonumber  \\
& & \left \{ e^{-\frac{2(\xi-k_0)^2}{\sigma^2}}e^{-\frac{\eta^2}{2\sigma^2}}\   S_{{\bf n_f},{\bf n_i}}(\xi+\frac{\eta}{2})\ S^{\star}_{{\bf n_f},{\bf n_i}}(\xi-\frac{\eta}{2})\ e^{-i\frac{\hbar}{m}\eta \xi\tau} \right \}  .
\label{poftau2}
\end{eqnarray}
Notice the much more intricate coupling between the integration variables $\xi$ and $\eta$ in the present expression for the delay-time probability.
The mean delay time during a specific transition
 $$\langle\tau\rangle_{{\bf n_f},{\bf n_i}}=\int_{-\infty}^{\infty} {\rm d}\tau \tau P_{{\bf n_f},{\bf n_i}}(\tau)$$
 can be computed from (\ref {poftau2}) in the same way as in (\ref {tauem}). Taking advantage of the condition $\sigma<k_0$  this results in
\begin{eqnarray}
\hspace{-25mm}  \langle\tau\rangle_{{\bf n_f},{\bf n_i}}=
 \ \frac{ m}{\hbar} \sqrt{\frac{2}{\pi \sigma^2}}\int_{-\infty}^{\infty}{\rm d}\xi   \xi^{-1}  e^{-\frac{2(\xi-k_0)^2}{\sigma^2}} i
  \frac{\partial\ }{\partial \eta}\left[ S_{{\bf n_f},{\bf n_i}}(\xi+\frac{\eta}{2})\ S^{\star}_{{\bf n_f},{\bf n_i}}(\xi-\frac{\eta}{2})\right ]_{\eta=0}   .
\end{eqnarray}
Since the $S$ matrix is a function of the energy $E$, one can write
\begin{equation}
 \hspace{-18mm} \langle\tau\rangle_{{\bf n_f},{\bf n_i}}=
\hbar \int_{-\infty}^{\infty}{\rm d}\xi  {\sqrt{\frac{2}{\pi \sigma^2}}   e^{-\frac{2(\xi-k_0)^2}{\sigma^2}}}
 \mathcal{I}{\it m}\left[ S_{{\bf n_f},{\bf n_i}}(E(\xi)) \frac{\partial\ }{\partial E}\ S^{ \dag}_{{\bf n_i},{\bf n_f}}(E(\xi) )\right ] .
\end {equation}
In the monochromatic limit $\sigma \rightarrow 0 $, the normalized Gaussian turns into a $\delta$ function and
\begin{equation}
 \langle \tau (E_0)_{{\bf n_f},{\bf n_i}}\rangle =
\hbar \
 \mathcal{I}{\it m}\left[ S_{{\bf n_f},{\bf n_i}}(E_0 ) \frac{\partial\ }{\partial E}\ S^{ \dag}_{{\bf n_i},{\bf n_f}}(E_0)\right ] \ .
\end {equation}
Averaging over the open initial channels and summing over the open final channels one gets
$$\langle \tau E _0\rangle=\frac{\hbar}{N_0} \mathcal{I}{\it m}\left[{\rm tr} \left (S(E_0)\frac{\partial\ }{\partial E}\ S^{ \dag}(E_0)\right )\right ]$$
where the trace is over the open channels exclusively. This coincides with
the standard form of Smith's version of Wigner's mean delay time \cite {Smith}.

The probability distribution (\ref {poft1}) can also be written  as an integral over the energy variables instead of the wave number variables.  Defining the mean energy $E=\frac{\hbar^2 }{2m}(k^2+k'^2)/2$ and the difference $\epsilon = \frac{\hbar^2 }{2m} (k^2-k'^2)$, one can obtain an approximate expression by neglecting terms of order $(\frac{\sigma}{k_0})^2$ in the exponent:
\begin{eqnarray}
\label{poft3}
\hspace{-20mm} P_{{\bf n_f},{\bf n_i}}(\tau)&=&\frac{1}{2\pi\hbar}\int_{-\infty}^{\infty} {\rm d}
\epsilon\ e^{-\frac{i}{\hbar}\epsilon\tau} e^{-\frac{1}{8}\frac{\epsilon^2}{\rho^2 E_0}} \times
  \\  \hspace{-20mm}
&\ &\left  \{\frac{1}{2}\sqrt{\frac{2}{\pi \rho^2}}
\int_{-\infty}^{\infty}\frac{{\rm d}E}{\sqrt{E}}e^{-2(\frac{\sqrt{E}-\sqrt{E_0}}{\rho })^2}\   S_{{\bf n_f},{\bf n_i}}(E+\frac{\epsilon}{2})S^{\star}_{{\bf n_f},{\bf n_i}}(E-\frac{\epsilon}{2})\right\}\nonumber
\end{eqnarray}
where $\rho=\frac{\sigma \hbar}{\sqrt{2m}}$ and the mean energy is  $E_0=E(k_0)$. In complete analogy to (\ref{poftem2prime}) the expression  in the curly brackets is the energy auto-correlation function of the matrix element $S_{{\bf n_f},{\bf n_i}}(E)$.

In the following paragraph the semi-classical approximation for the scattering matrix will be used to show that in the semi-classical limit, (\ref{poft3}) tends to the  classical  delay-times distribution.
The semi-classical approximation for the scattering matrix is built upon classical trajectories which generalize  the ones defined in the introduction section to include inelastic scattering. Here, the classical trajectories start in the far past $(t=-T)$ far away from the interaction domain,  with relative momentum  ${\bf p}_i$, and with the internal action variables ${\bf I}_i= \hbar {\bf n}_i$ corresponding to the quantum numbers of the initial state. Far in the future, $(t=+T)$ the relative momentum is ${\bf p}_f$ and the action variables for the internal dynamics are ${\bf I}_f= \hbar {\bf n}_f$ corresponding to the final state quantum numbers. Energy conservation demands, $E=\frac{|{\bf p}_f|^2}{2m} + \epsilon_{{\bf n}_f} \ = \frac{|{\bf p}_i|^2}{2m} + \epsilon_{{\bf n}_i}.$ Following the convention introduced previously, the directional degrees of freedom $\Omega = (\theta, \phi)$ and their conjugate momenta $(l_{\theta},l_{\phi})$ are concatenated to the list of internal action-angle variables. With $r$ the radial distance and $p_r$ its conjugate momentum, the reduced action reads now
\begin{eqnarray}
\label{genaction}
\hspace{-5mm} \Phi({\bf n}_f,{\bf n}_i;E)&=&-\int_{-T}^{+T}{\rm d}t\ \left [   r(t)\dot{p}_r (t)    +  ({\boldsymbol \varphi}(t)\cdot {\bf \dot{I}}(t)  )\right ]\ .
 \end{eqnarray}
The semiclassical approximation \cite{Miller,leshouches} for the $S$ matrix reads:
\begin{equation}
S_{{\bf n}_f,{\bf n}_i}(E)\ \approx\ \sum_{\alpha}[P^{(\alpha)}_{{\bf n}_f,{\bf n}_i}(E)]^{\frac{1}{2}}
e^{\frac{i}{\hbar}\Phi^{(\alpha)}({\bf n}_f,{\bf n}_i;E)-i\frac{\pi}{2}\nu^{(\alpha)}}\ ,
\label{semiclass}
\end{equation}
Where, the different classical trajectories which are compatible with the transition ${\bf n}_i\rightarrow {\bf n}_f$ at energy $E$ are distinguished by the index $\alpha$. Each trajectory contributes an amplitude which is the square root of the  classical delay probability
$P^{(\alpha)}_{{\bf n}_f,{\bf n}_i}(E)$ and a phase factor $e^{-i\frac{\pi}{2}\nu^{(\alpha)}}$ where $\nu^{(\alpha)}$ is the Maslov index. Both quantities are derived from the Hessian $ W^{(\alpha)}_{r,s}=\frac{\partial^2 \Phi({\bf n}_f,{\bf n}_i;E)}{\partial n_r,\partial n_s}$ computed at the trajectory $\alpha$. Then, $P^{(\alpha)}_{{\bf n}_f,{\bf n}_i}(E)=|\det W^{(\alpha))}|$ and $\nu^{(\alpha)}$ is the number of its negative eigenvalues (Morse index).

Substituting (\ref {semiclass}) in (\ref {poft3}) and using the fact that $\frac{\partial\Phi^{(\alpha)}({\bf n}_f,{\bf n}_i;E)}{\partial E} = \tau^{(\alpha)}_{{\bf n}_f,{\bf n}_i}$ we can approximate the term in the curly bracket by
\begin{equation}
\hspace{-2  mm} \left \langle
\sum_{\alpha,\beta}[P^{(\alpha)}P^{(\beta)}]^{\frac{1}{2}}
e^{\frac{i}{\hbar}[\Phi^{(\alpha)}-\Phi^{(\beta)}]-i\frac{\pi}{2}\delta\nu^{(\alpha,  \beta)}}
\ e^{ \frac{i}{2\hbar}\epsilon [\tau^{(\alpha)}+\tau^{(\beta)}]}
\right \rangle_{E}
\label{semiclass1}
\end{equation}
To simplify the notation the suffices ${\bf n}_f,{\bf n}_i$ were suppressed,  every term which appears in square brackets is a function of  $E$, and $\delta\nu^{(\alpha,  \beta)}$ denotes the difference between the Maslov indices. The triangular brackets stand for integrating with respect to $E$ with the Gaussian weight as in (\ref {poft3}). The main contribution to (\ref {semiclass1}) comes from the diagonal sum  with $\alpha=\beta$. This partial sum will be discussed first, and the discussion of the non-diagonal will follow.
Substituting in the diagonal sum in (\ref {poft3}) and integrating over $\epsilon$, one gets
\begin{equation}
P^{diag}_{{\bf n_f},{\bf n_i}}(\tau) \approx \left \langle
\sum_{\alpha} P^{(\alpha)}(E)\frac{1}{\sqrt{\pi \Delta^2}}e^{-(\frac{\tau-\tau^{(\alpha)} (E)}{\Delta})^2}
\right \rangle_{E}
\label{semiclass2}
\end{equation}
where $\Delta= \frac{\hbar}{E_0} \frac{\sqrt{2}k_0}{\sigma}$ is the time resolution. It equals the uncertainty imposed by Heisenberg times a factor which expresses the worsening of the resolution due to the restriction of the band width $\sigma$ to be smaller than the carrier wave number $k_0$. Clearly, $\Delta$  is a small number in the semi-classical domain, and hence the classical limit emerges as
\begin{eqnarray}
\label{classdiag}
P^{cl}_{{\bf n_f},{\bf n_i}}(\tau) &\approx& \left \langle
\sum_{\alpha} P^{(\alpha)}(E)\delta(\tau-\tau^{(\alpha)} (E))
\right \rangle_{E}  \ .
\end{eqnarray}

Expression (\ref {classdiag}) has a simple intuitive interpretation. For every $E$ in the wave packet there are several trajectories which have the same delay-time. The classical probability carried by these trajectories is   $\sum_{\alpha} P^{(\alpha)}(E)$. The total probability  is obtained after averaging over the continuous $E$ distribution as  in (\ref {poft3}) which will be denoted by $W(E)$ to simplify the notation. The delay time probability  can then be written in the form
 \begin{equation}
 \sum_{\alpha} P^{(\alpha)}(E^{(\alpha)}(\tau))W(E^{(\alpha)}(\tau))\left |\left [ \frac{\partial E^{(\alpha)}(t)}{\partial  t } \right ]_{t=\tau}\right |\ .
 \end{equation}
 Here $E^{(\alpha)}(\tau)$ is the (local) inverse of $\tau^{(\alpha)}(E)$ and in performing the integral one has to assume that no singularities of the classical trajectories due to e.g., caustics occur in the energy window which supports $W(E)$.

Coming back to the non-diagonal sum in (\ref {semiclass1}), one can compute the contribution from pairs of different trajectories using the same  way as above to get a correction of the form
\begin{equation}
\hspace{-10mm} \left \langle
\sum_{\alpha\ne \beta}[P^{(\alpha)}P^{(\beta)}]^{\frac{1}{2}}
e^{\frac{i}{\hbar} [\Phi^{(\alpha)}-\Phi^{(\beta)}] -i\frac{\pi}{2} \delta\nu^{(\alpha,  \beta)} }
\delta (\tau-\frac{[\tau^{(\alpha)}+\tau^{(\beta) }]}{2})
\right \rangle_{E}\ .
\label{semiclass3}
\end{equation}

Expressions of this type were discussed previously in various contexts \cite{argaman,berkolaiko,sieber,heusler} and in particular in \cite {jack, Saar}. In the present context (\ref{semiclass3}) can be considered as a correlation function of (reduced) actions of pairs of trajectories having the same energy, but with delay times whose average is $\tau$. The main contributions come from pairs of trajectories with action differences of order $\hbar$. Such pairs become more rare as one approaches the classical limit $\hbar\rightarrow 0$,
unless there are exact degeneracies due to symmetries in the scattering object. Thus, for generic systems this term  does not contribute to the classical expression. However, in scattering on graphs, exact degeneracies do occur, and such terms are important. They will be discussed  in the following paper of this series.

\subsection {An illustration and Conclusion}

To illustrate the ideas and results developed above, consider a scattering problem in the plane, where waves are scattered by a group of  $N\ge 3$  perfectly reflecting circular mirrors of unit radius. The discs are distributed in the plane without overlaps,  and the smallest circle which encloses them is denoted by $C$   \cite {Gaspard, Wirzba}.  Denote the positions of the circle centers by ${\boldsymbol \rho}_i $, and the unit vectors from a center ${\boldsymbol \rho}_i $ to points on its circle  by ${\bf r}_i $.

To compute the scattering matrix in the semi-classical limit, one should identify the classical trajectories which start in the direction ${\bf \Omega}_{i}$  outside the circle $C$ and end after scattering  outside $C$ in the direction  $\bf{\Omega}_{f}$. Denote the position of points on the reflecting circles by ${\bf R}_i= {\boldsymbol \rho}_i+{\bf r}_i $. The classical scattering trajectories are encoded by a list of circles ${\bf a}=(a_1,\cdots,a_M)$, $a_m\in \{1,\cdots,N\}$ from which they scatter before escaping. Naturally, $a_m\ne a_{m+1}$. Given a code ${\bf a}$ consider a polygonal path which goes through the points ${\bf R}_{a_m}$ and denote its length by
\begin{equation}
L_{\bf a} = \sum_{m=2}^M |{\bf R}_{a_m}-{\bf R}_{a_{m-1}}|\ .
\end{equation}
$L_{{\bf a}}$ is a function of the points ${\bf r}_{a_m}$ on the reflecting circles. The  classical scattering trajectories of a certain code are the extrema of $L_{{\bf a}}$, which guarantee that the reflections on the reflecting circles are specular. They must also conform with the following constraints:

\noindent \emph{i}. The incoming and outgoing segments of the trajectory satisfy specular reflection conditions
\begin{equation} \hspace{-10mm}
({\bf \Omega}_i\cdot {\bf r}_{a_1})=-({\bf r}_{a_1}\cdot {\bf {\hat R}}_{{a_1}{a_2}}) \ \  ;\ \ ({\bf \Omega}_f\cdot {\bf r}_{a_M})=-({\bf r}_{a_M}\cdot {\bf {\hat R}}_{{a_{M-1}}{a_M}})
\end{equation}
where ${\bf {\hat  R}}_{a_{m} a_{m+1}}$ denotes a unit vector in the direction of $({\bf  {R}}_{ a_{m+1}}-{\bf { R}}_{a_{m}})$.

\noindent \emph{ii}. All the segments of an extremal trajectory, (including the incoming and outgoing ones) do not intersect the interiors of any of the circles.

To any classical scattering trajectory one associates a probability amplitude which takes into account the fact that the scattering from a circular reflector is not uniform in the deflection angle:
\begin{equation}
A[{\bf a};\Omega_i,\Omega_f] = \frac{(-1)^M}{2^{\frac{M}{2}}}\left [\prod_{m=1}^M  \sqrt{ (1-({\bf {r}}_{a_{m}}\times {\bf{\hat  R}}_{a_{m} a_{m+1}})^2)}  \right ] ^{\frac{1}{2}},
\label{Ascat}
\end{equation}
where, ${\bf{\hat  R}}_{a_{0} a_{1}}={\bf \Omega}_i$ and ${\bf{\hat  R}}_{a_{M} a_{M+1}}={\bf \Omega}_f$. The phase factor is due to the Dirichlet boundary condition on the reflecting surfaces.   It should be noted that not for every code word there exists a classical trajectory which satisfies the boundary conditions and the constraints above. However, the theory of chaotic billiards guarantees that the codes are unique, and the number of trajectories which are reflecting $M$ times grows exponentially with $M$.

Using this information and (\ref {semiclass}), the semi-classical expression for the $S(E)$ matrix element is
\begin{equation}
S_{\Omega_f,\Omega_i}(k)\ \approx\ \sum_{\bf a} A[{\bf a};\Omega_i,\Omega_f]
e^{ i k L[{\bf a};\Omega_i,\Omega_f] }\ ,
\label{semiclasscir}
\end{equation}
and the sum goes over all the codes of allowed trajectories with $M \ge 1$. The dimensionless actions $\frac{1}{\hbar}\Phi$ are replace by $k L[{\bf a};\Omega_i,\Omega_f]]$, the Maslov indices vanish, and they are replaced by the reflection phase included in $A$ (\ref{Ascat}). Notice that the only energy dependence in this expression is in the phase factor. Thus, the Fourier transform of  auto-correlation function in the semi-classical approximation is
\begin{eqnarray}
P_{\Omega_i,\Omega_f}(s)&=&
\frac{1}{2\pi} \int {\rm d}\eta e^{-i\eta s}\left \langle S_{\Omega_f,\Omega_i}(k+\frac{\eta}{2})S_{\Omega_f,\Omega_i}(k-\frac{\eta}{2})
\right\rangle_k \nonumber \\
&\approx& \sum_{\bf a} |A[{\bf a};\Omega_i,\Omega_f]|^2\delta(s-L[{\bf a};\Omega_i,\Omega_f])
\end{eqnarray}
Excluding from the sum the trajectories which reflect from a single disc, the remaining sum is the classical probability that a trajectory impinges on the set of scatterers from the direction $\Omega_i$ remains confined to the interior for a length $s=v_0t$  before it escapes in the direction $\Omega_f$. Since the reflections induce chaotic dynamics, the integral over the final directions averaged over the incoming directions gives the probability that a trajectory staring in the interior would travel a distance $s$ before escaping. This quantity is known from the theory of classical chaotic scattering to decay exponentially with a decay constant $\gamma$  which can be expressed in terms of the Lyapunov exponent $\lambda$ and the Hausdorff dimension $d_H $ of the set of initial angular momenta which correspond to trapped trajectories: $\gamma = (1-d_H)\lambda$ \cite{leshouches}.  Thus,
\begin{eqnarray}
P_(s) =
\frac{1}{4\pi^2} \int {\rm d}\eta e^{-i\eta s}\left \langle {\rm tr }[ S (k+\frac{\eta}{2})S^\dag(k-\frac{\eta}{2})]
\right\rangle_k
 \approx  \gamma e^{-\gamma s}
\end{eqnarray}
 expresses the probability distribution function as the Fourier transform of the $S$ matrix autocorrelation function.

 In conclusion, the present paper expanded and generalizes the idea first formulated in \cite{Lyuboshits,blumel,leshouches} that the quantum delay-time distribution can be expressed as the form-factor of the scattering matrix. In particular, the important role played by the envelope function in determining the observed delay-time probability distribution function was emphasized. The approach of the quantum mechanical distribution to the classical limit was  discussed, and Wigner's result for the mean delay-time was shown to hold in the monochromatic limit.

\section{Acknowledgements}
The author is indebted to Professor Saar Rahav for his careful  reading of an earlier version of this manuscript, making valuable suggestions and comments, and in particular,  bringing to my attention the paper by  V.L. Lyuboshits. I appreciate highly and thank Drs Christopher Joyer and Sven Gnutzmann for critical comments, suggestions and corrections.   Professor Steve Anlage is acknowledged for directing me to the subject by sharing with me his preliminary experimental results. Thanks are also due to Professor Nirit Dudovich for introducing me to ultra-fast optics.

\end{document}